\documentclass[a4paper, 11pt]{article}

\usepackage[affil-it]{authblk} 
\usepackage{etoolbox}
\usepackage{lmodern}
\usepackage{subcaption}
\makeatletter
\patchcmd{\@maketitle}{\LARGE \@title}{\fontsize{16}{16}\selectfont\@title}{}{}
\makeatother

\renewenvironment{abstract}
 {\small
  \begin{center}
  \bfseries \abstractname\vspace{-.5em}\vspace{0pt}
  \end{center}
  \list{}{%
    \setlength{\leftmargin}{5mm}
    \setlength{\rightmargin}{\leftmargin}%
  }%
  \item\relax}
 {\endlist}

\usepackage{lineno}
\usepackage{amssymb}
\usepackage{amssymb,latexsym, graphicx, amsfonts, caption, color, verbatim}
\usepackage[fleqn]{amsmath, mathtools}
\usepackage{graphics}
\usepackage{graphicx}
\usepackage{array}

\usepackage{subcaption}

\title{Impact of the adjustment of stratification factors on time-to-event analyses}
\author{\vspace{0.2in} Madan G. Kundu$^1$ and Shoubhik Mondal$^2$\\
 $^1$AbbVie Inc, IL, USA; $^2$Shoubhik Mondal, Boehringer Ingelheim, CT, USA }
\date{}
\begin{document}

\maketitle


\begin{abstract}
In a stratified clinical trial design with time to event end points, stratification factors are often accounted for the log-rank test and  the Cox regression analyses. In this work, we have evaluated the  impact of inclusion of stratification factors on the power of the stratified log-rank test and have compared the bias and standard error in HR estimate  between multivariate and stratified Cox regression analyses through simulation. Results from our investigation suggests that both failing to consider stratification factor in presence of their prognostic effect and stratification with smaller number of events may substantially reduce the power of the log-rank test. Further, the HR estimate from the  multivariate Cox analysis  is more accurate and precise compared to the stratified Cox analysis.  Our findings point towards the necessity of evaluating the impact of stratification factors on the time to event analyses at the time of study design which is presently not a norm.
\end{abstract}


\vspace{8mm}
In clinical trials  comparing time to event endpoints such as progression free survival (PFS) or overall survival (OS) between two treatment arms, some baseline covariates can influence the treatment benefit either due to their prognostic effect or their association with treatment responsiveness \cite{kernan1999stratified, tu2000adjustment, iche9}. To address this, often stratified randomization is employed considering these baseline covariates as stratification factors. The regulatory guidelines \cite{iche9, committee2004committee} suggest to include the stratification factors on which randomization has been stratified  later in the analysis stage and accounting for them in the analysis stage are seen as complementary to stratified randomization. In compliance with these guidelines, hypothesis testing are generally carried out in clinical trials using stratified log-rank test whereas treatment effects are quantified using hazards ratio (HR) from stratified or multivariate Cox regression model. Given the strengths and weaknesses of both stratified and multivariate analyses and the mixed practice, the purpose of this work is to evaluate the impact of inclusion of stratification factors on the power of stratified log-rank test and to compare the bias and standard error in HR estimate  between multivariate and stratified Cox regression analyses through simulations.\\

To put it simply, under stratified log-rank test, test statistics are derived separately within each stratum and then added together instead of deriving test statistic based on the entire population as done in standard unstratified log-rank test. Similarly, in the stratified Cox regression model,  HR is calculated within each stratum  and then combine these HR to obtain a global HR \cite{kalbfleisch2011statistical}. An alternative approach to the stratified Cox regression model is to use multivariate Cox regression model where stratification factors are passed as covariates in addition to the treatment, as originally proposed by Cox \cite{cox1972regression}. In our observation, majority of recently published pivotal clinical trials (e.g., see \cite{kang2020early, slamon2020overal, de2019cabazitaxel, paz2017afatinib}) have reported HR based on stratified Cox regression analysis  whereas the use of multivariate Cox model has been very limited in recent times \cite{suntharalingam2017effect, ng2019effect}. In some cases, HR from both the models are reported \cite{raymond2011sunitinib}. The use of multivariate Cox model is limited to exploratory analysis \cite{touchefeu2019prognostic} or the clinical trials with small sample size \cite{mohnike2019radioablation, wirsching2018bevacizumab}. 
\\

Rational for using stratified log-rank test and stratified Cox regression analysis stems from the need to control the prognostic effects of stratification variables, a common rational to any other stratified analysis as well.  However, stratification may cause loss of power \cite{akazawa1997power, feng2010power}and estimate of HR may not be stable \cite{tu2000adjustment}, especially when the number of strata is very large relative to the sample size. On contrary, although multivariate Cox model leads to more stable estimate, it's use has been criticized as it may lead to biased estimate of HR due to the strong assumption that stratification factors have a multiplicative effect on the hazard rate \cite{hill1981asymptotic}. Akazawa et. al. \cite{akazawa1997power} carried out simulations to compare power between stratified log-rank test and the standard log-rank test primarily under two scenarios: first, there is no prognosis effect of stratification factors and second, the stratification factors affect the true HR. Ding and Sinha \cite{dingevaluation} compared power between these stratified and multivariate Cox models, but they did not report the bias or standard error in HR estimation. We have carried out simulation to evaluate the properties (i.e., power, bias and standard error) of stratified and multivariate analyses under a more  realistic clinical  trial scenario with prognostic effect of stratification factors limited to the baseline hazards within each strata but  not on the treatment effect measured by treatment to control HR. Precisely, we have considered a randomized two arms clinical trial comparing treatment arm with control arm with 14 months of accrual, 12 strata based on 3 stratification factors: $X_1$ (with 2 levels), $X_2$ (with 3 levels) and $X_3$ (with 2 levels). Our simulation was carried out to reflect following three distinct scenarios:

\vspace{-2mm}
\begin{itemize} \setlength\itemsep{0em}
\item{Scenario 1 (No prognostic effect):} No effects of stratification factors.
\item{Scenario 2 (Multivariate Cox model):} Prognostic effect of stratification factors on baseline hazards (but not the treatment to control HR) in consistent with multivariate Cox model.
\item{Scenario 3 (Stratified Cox model):} Prognostic effect of stratification factors on baseline hazards (but not the treatment to control HR) that is not consistent with multivariate Cox model. 
\end{itemize}

\vspace{-2mm}

Event times for the control arm were generated as follows: (a) under ``no prognostic effect'' model, from a hazards function with median time as 16 months, (b) under ``multivariate Cox model'', from hazard function $\lambda(t|x)=\lambda(t)\exp(\beta_{x1}x_1+\beta_{x21}x_{21}+\beta_{x22}x_{22}+\beta_{x3}x_3)$ for stratum $X_1=x_1, X_2=(x_{21}, x_{22}), X_3=x_3$ with $X_{21}$ and $X_{22}$ being the indicator variables for $X_2$, $\lambda(t)=log(2)/16$, $\exp(\beta_{x1})=0.5$, $\exp(\beta_{x21})=0.75$,  $\exp(\beta_{x22})=1.25$ and  $\exp(\beta_{x3})=0.75$ translating strata specific median time as 16,  21.3, 21.3, 28.4, 12.8, 17.1, 32, 42.7, 42.7, 56.9, 25.6 and 34.1 months, respectively, and (c) under ``stratified Cox model'', from stratum specific 12 distinct baseline hazard functions (not consistent with  multiplicative nature of a multivariate Cox model) with corresponding median as  16, 16, 16, 16, 16, 16, 50, 50, 50, 50, 50 and 50 months.  Within each stratum, event times under both the treatment arm and the control arm were generated according to exponential distribution with  hazard rate for the  treatment arm  obtained by multiplying control hazard rate with  the  true HR (0.50 to 0.75). Further, for each patients, enrollment times were generated randomly from uniform distribution and assignments to the treatment arms were made according to the Bernoulli distribution with equal probabilities  reflecting 1:1 randomization. Multivariate Cox model was fitted  including treatment indicator variable and $X_1$, $X_2$ and $X_3$ as covariates without any interaction term as suggested in \cite{committee2004committee}. Stratified Cox model was fitted including $X_1$, $X_2$ and $X_3$ as stratification factors and treatment as the only covariate. Simulations  were carried out with  minimum events (D) sufficient for 80\% power with  corresponding sample sizes as $D/0.70$. Strata sizes were varied from balanced strata size to unabalanced strata sizes towards strata with better prognosis or poor prognosis. Simulation codes are included as supplementary material to this paper.\\ 

\begin{table}
\begin{center}
\caption{Performance of Cox models (multivariate and stratified) and  log-rank tests (stratified and unstratified) in estimating and testing HR (treatment to control) based on 10,000 simulations}
\label{tab1}
\resizebox{\textwidth}{!}{%
\begin{tabular}
{l|l|c|r|r|r|r|r|r|r|r|r|r|r|r|r}
\hline
&&& \multicolumn{3}{c|}{Average Bias} & \multicolumn{3}{c|}{Average SE} & \multicolumn{3}{c|}{Average MSE} & \multicolumn{4}{c}{}\\ 
&&& \multicolumn{3}{c|}{in HR estimate} & \multicolumn{3}{c|}{in HR estimate} & \multicolumn{3}{c|}{in HR estimate} & \multicolumn{4}{c}{Power(\%)}\\  \cline{4-16}
& True& Eve-& & Mult  & Strat  & & Mult  & Strat& & Mult  & Strat  &  & Strat  & Mult  & Strat  \\ 
Strata size (ratio)&HR &nts & Cox& Cox & Cox & Cox & Cox & Cox &Cox & Cox & Cox & LR & LR & Cox & Cox \\ 
\hline
\multicolumn{16}{l}{}\\ 
\multicolumn{16}{l}{\textbf{Simulation scenario 1:} no prognostic effect of stratification factors }\\ 
\hline
Balanced strata size &  0.5 &  66 & 0.015 & 0.007 & 0.016 & 0.255 & 0.262 & 0.294 & 0.018 & 0.019 & 0.024 & 79.1 & 68.6 & 78.1 & 67.8 \\ 
  Ratio of size: & 0.55 &  88 & 0.014 & 0.009 & 0.015 & 0.219 & 0.223 & 0.244 & 0.016 & 0.016 &  0.02 & 78.7 & 70.6 & 78.2 & 70.2 \\ 
  1:1:1:1:1:1:1:1:1:1:1:1 &  0.6 & 120 & 0.009 & 0.006 &  0.01 & 0.186 & 0.188 &   0.2 & 0.013 & 0.014 & 0.015 &   79 & 73.9 & 78.4 & 73.5 \\ 
   & 0.65 & 170 &  0.01 & 0.008 &  0.01 & 0.156 & 0.157 & 0.164 & 0.011 & 0.011 & 0.012 & 78.8 & 74.6 & 78.3 & 74.5 \\ 
   &  0.7 & 248 & 0.006 & 0.004 & 0.006 & 0.128 & 0.129 & 0.133 & 0.008 & 0.008 & 0.009 & 79.8 & 77.2 & 79.6 & 77.1 \\ 
   & 0.75 & 380 & 0.004 & 0.003 & 0.004 & 0.103 & 0.104 & 0.106 & 0.006 & 0.006 & 0.006 & 79.4 & 77.5 &   79 & 77.4 \\
 \hline
\multicolumn{16}{l}{}\\ 
 \multicolumn{16}{l}{\textbf{Simulation scenario 2:} Multivariate Cox model}\\ 
\hline
  Balanced strata size &  0.5 &  66 & 0.049 & 0.016 & 0.019 & 0.254 & 0.262 & 0.297 & 0.022 &  0.02 & 0.024 & 71.2 & 67.1 & 75.9 & 66.6 \\ 
  Ratio of size: & 0.55 &  88 & 0.044 & 0.012 & 0.013 & 0.218 & 0.224 & 0.247 & 0.019 & 0.017 &  0.02 & 71.3 & 69.7 & 76.6 & 69.2 \\ 
  1:1:1:1:1:1:1:1:1:1:1:1 &  0.6 & 120 & 0.036 & 0.012 & 0.011 & 0.186 & 0.189 & 0.201 & 0.016 & 0.014 & 0.016 & 72.8 & 73.3 & 77.1 &   73 \\ 
   & 0.65 & 170 & 0.031 &  0.01 & 0.008 & 0.155 & 0.157 & 0.164 & 0.012 & 0.011 & 0.012 & 73.1 & 75.6 & 77.7 & 75.3 \\ 
   &  0.7 & 248 & 0.028 & 0.009 & 0.006 & 0.128 & 0.129 & 0.133 &  0.01 & 0.009 & 0.009 & 73.1 & 76.8 & 78.3 & 76.6 \\ 
   & 0.75 & 380 & 0.022 & 0.007 & 0.004 & 0.103 & 0.104 & 0.106 & 0.007 & 0.006 & 0.006 & 72.7 &   78 & 78.1 & 77.9 \\
 \hline
  Unbalanced (more  &  0.5 &  66 & 0.035 & 0.015 & 0.018 & 0.255 & 0.262 & 0.292 &  0.02 &  0.02 & 0.024 & 75.3 & 68.7 & 76.7 &   68 \\ 
  subjects in strata   & 0.55 &  88 & 0.031 & 0.014 & 0.014 & 0.219 & 0.224 & 0.245 & 0.018 & 0.017 &  0.02 & 74.8 &   70 & 76.8 & 69.6 \\ 
  with better prognosis) &  0.6 & 120 & 0.024 & 0.011 & 0.009 & 0.186 & 0.189 & 0.201 & 0.014 & 0.014 & 0.015 &   76 & 74.3 & 78.2 &   74 \\ 
  Ratio of size: & 0.65 & 170 & 0.023 & 0.012 & 0.009 & 0.155 & 0.157 & 0.165 & 0.012 & 0.011 & 0.012 & 75.2 & 74.7 & 77.4 & 74.4 \\ 
  1:1:1:1:1:1:7:7:7:7:7:7 &  0.7 & 248 & 0.018 & 0.008 & 0.005 & 0.128 & 0.129 & 0.134 & 0.009 & 0.008 & 0.009 & 76.4 & 76.9 & 78.7 & 76.7 \\ 
   & 0.75 & 380 & 0.015 & 0.007 & 0.003 & 0.103 & 0.104 & 0.106 & 0.006 & 0.006 & 0.006 & 75.8 & 77.7 & 78.7 & 77.5 \\ 
 \hline
  Unbalanced (more  &  0.5 &  66 & 0.035 & 0.013 & 0.018 & 0.255 & 0.262 & 0.284 &  0.02 &  0.02 & 0.023 & 74.7 & 70.6 & 77.2 &   70 \\ 
 subjects  in strata   & 0.55 &  88 & 0.032 & 0.013 & 0.013 & 0.218 & 0.223 & 0.238 & 0.018 & 0.017 & 0.019 & 74.3 & 72.1 &   77 & 71.6 \\ 
  with poor prognosis) &  0.6 & 120 & 0.021 & 0.009 & 0.009 & 0.186 & 0.188 & 0.196 & 0.014 & 0.014 & 0.014 & 76.2 & 75.3 &   78 & 75.1 \\ 
  Ratio of size: & 0.65 & 170 & 0.019 & 0.007 & 0.007 & 0.155 & 0.157 & 0.162 & 0.011 & 0.011 & 0.012 & 76.1 & 77.2 & 78.9 & 76.9 \\ 
  7:7:7:7:7:7:1:1:1:1:1:1 &  0.7 & 248 & 0.017 & 0.007 & 0.005 & 0.128 & 0.129 & 0.132 & 0.009 & 0.009 & 0.009 & 75.7 & 77.2 & 78.4 &   77 \\ 
   & 0.75 & 380 & 0.013 & 0.004 & 0.002 & 0.103 & 0.104 & 0.105 & 0.006 & 0.006 & 0.006 & 76.3 & 79.1 & 78.5 &   79 \\
  \hline
  \multicolumn{16}{l}{}\\ 
 \multicolumn{16}{l}{\textbf{Simulation scenario 3:} Stratified Cox model}\\ 
\hline
  Balanced strata size &  0.5 &  66 & 0.071 & 0.006 & 0.017 & 0.253 & 0.264 & 0.299 & 0.026 &  0.02 & 0.025 & 65.7 & 66.6 & 77.8 & 65.9 \\ 
  Ratio of size: & 0.55 &  88 & 0.066 & 0.005 & 0.012 & 0.218 & 0.225 & 0.249 & 0.023 & 0.017 &  0.02 & 65.4 & 69.7 & 78.2 & 69.1 \\ 
  1:1:1:1:1:1:1:1:1:1:1:1 &  0.6 & 120 & 0.056 & 0.004 & 0.009 & 0.185 & 0.189 & 0.202 & 0.018 & 0.014 & 0.015 & 66.4 & 73.3 & 79.4 & 72.9 \\ 
   & 0.65 & 170 & 0.049 & 0.003 & 0.006 & 0.155 & 0.157 & 0.165 & 0.014 & 0.011 & 0.012 &   67 & 75.5 & 79.8 & 75.3 \\ 
   &  0.7 & 248 & 0.045 & 0.004 & 0.007 & 0.128 & 0.129 & 0.134 & 0.011 & 0.009 & 0.009 & 65.9 & 76.1 &   79 & 75.9 \\ 
   & 0.75 & 380 & 0.038 & 0.003 & 0.004 & 0.103 & 0.104 & 0.106 & 0.008 & 0.006 & 0.006 & 65.5 & 77.7 & 79.8 & 77.6 \\ 
 \hline
  Unbalanced (more  &  0.5 &  66 & 0.037 & 0.006 & 0.016 & 0.255 & 0.264 & 0.294 & 0.021 &  0.02 & 0.024 & 74.4 & 68.6 &   78 & 67.9 \\ 
  subjects in strata   & 0.55 &  88 & 0.038 & 0.008 & 0.016 & 0.219 & 0.225 & 0.247 & 0.018 & 0.016 &  0.02 &   73 & 69.4 & 77.7 &   69 \\ 
  with better prognosis) &  0.6 & 120 &  0.03 & 0.004 & 0.009 & 0.186 & 0.189 & 0.202 & 0.015 & 0.014 & 0.016 & 73.5 & 73.3 & 78.7 & 72.9 \\ 
  Ratio of size: & 0.65 & 170 & 0.026 & 0.006 & 0.008 & 0.155 & 0.157 & 0.166 & 0.012 & 0.011 & 0.012 & 74.3 & 74.7 & 78.8 & 74.4 \\ 
  1:1:1:1:1:1:7:7:7:7:7:7 &  0.7 & 248 & 0.023 & 0.004 & 0.006 & 0.128 & 0.129 & 0.134 & 0.009 & 0.008 & 0.009 & 74.5 & 76.4 & 79.9 & 76.2 \\ 
   & 0.75 & 380 & 0.016 & 0.002 & 0.003 & 0.103 & 0.104 & 0.107 & 0.007 & 0.006 & 0.006 & 75.2 & 77.9 & 80.2 & 77.7 \\ 
 \hline   
  Unbalanced (more  &  0.5 &  66 & 0.032 & 0.004 & 0.013 & 0.255 & 0.261 & 0.282 &  0.02 & 0.019 & 0.022 & 75.7 & 72.7 &   79 & 72.1 \\ 
  subjects in strata   & 0.55 &  88 & 0.034 & 0.008 & 0.014 & 0.218 & 0.223 & 0.237 & 0.018 & 0.016 & 0.019 & 73.6 &   73 & 78.3 & 72.5 \\ 
  with poor prognosis) &  0.6 & 120 & 0.023 & 0.007 & 0.011 & 0.186 & 0.188 & 0.195 & 0.014 & 0.014 & 0.015 & 75.8 & 75.7 & 78.3 & 75.5 \\ 
  Ratio of size: & 0.65 & 170 & 0.019 & 0.004 & 0.006 & 0.155 & 0.157 & 0.161 & 0.011 & 0.011 & 0.012 & 76.6 & 77.5 & 79.8 & 77.4 \\ 
  7:7:7:7:7:7:1:1:1:1:1:1 &  0.7 & 248 & 0.018 & 0.004 & 0.005 & 0.128 & 0.129 & 0.131 & 0.009 & 0.008 & 0.009 & 75.3 & 78.3 & 79.6 & 78.2 \\ 
   & 0.75 & 380 & 0.014 & 0.002 & 0.004 & 0.103 & 0.104 & 0.105 & 0.006 & 0.006 & 0.006 & 76.5 & 79.3 & 80.4 & 79.2 \\ 
 \hline
\multicolumn{16}{l}{-Mult: Multivariate; Strat: Stratified; Unstrat: Unstratified; LR: Log-rank test; SE: Standard error; HR: treatment }\\
\multicolumn{16}{l}{ arm to control arm hazards ratio}\\
\multicolumn{16}{l}{-Events (D) were set at required events for 80\% power with 2.5\% type-I error rate.}\\
\hline
\end{tabular}}
\end{center}
\end{table}

Table~\ref{tab1} summarize the results based on 10,000 simulations under all three scenarios for multivariate Cox analysis and stratified analysis.   Results from unstratified analyses which includes simple Cox regression and standard log-rank test without considering any stratification factors are also provided. In presence of prognostic effects of stratification factors (see Scenario 2 and 3), the  power from unstratified analyses are  consistently smaller than the power of stratified and multivariate analyses, largely driven by larger bias towards null in unstratified analyses. The power of unstratified log-rank test remains substantially smaller than the nominal 80\% power even in studies with larger number of events. This finding together with the power loss with unstratified log-rank test in presence of prognostic effects on HR reported by Akazawa et al. \cite{akazawa1997power} underlines the need to account for  stratification factors in time to events analyses. The performance of unstratified analyses improves with unbalance in strata size as this increases degree of homogeneity in overall sample compared to the sample with balanced strata size in which case degree of heterogeneity is at its maximum.\\

Under all the three scenarios (Scenario 1 - 3), the power of stratified log-rank test is consistently smaller than nominal power when number of events is small, but it gradually catches up to nominal 80\% power with the increase in the number of required events. This is consistent with the mathematical argument that the power of log-rank test is adversely affected by extreme value for the ratio of the number of patients at risk in the two treatment groups and such extreme values are more frequent in case of stratification \cite{akazawa1997power}. In comparison with standard Cox regression analysis, the HR estimate from stratified Cox regression analysis has greater standard error in consistent with  analytical proof provided by Feng et al.  \cite{feng2010power}. In relative assessment between stratified and multivariate analyses, the HR estimate from multivariate Cox model is less biased  and more precise (i.e., less standard error) than that of stratified Cox model under all scenario including when the data is generated under ``stratified Cox model''.  Due to greater bias and higher standard error, stratified Cox analysis has less power compared to multivariate Cox analysis, as observed previously \cite{dingevaluation}, implying that 95\% confidence interval for HR estimated from stratified Cox analysis is more likely to include $1$.  The difference in performance between stratified and multivariate analyses gradually fades away when study is designed with higher number of events. Our findings suggest that for studies with smaller number events, multivariate analysis would be prudent choice; however, as the number of events increases to 380, both the stratified and multivariate seem to be equally efficient.\\

Results from our investigation suggests that both failing to consider stratification factor in presence of their prognostic effect and stratification with smaller number of events may substantially reduce the power of log-rank test and at times this power loss exceeds 10\%. Noting that stratified log-rank test is often more obvious choice and recommended in regulatory guidelines, one must be very considerate about the power loss due to stratification when designing studies with smaller number of events. Presently, studies with stratified log-rank test are regularly  designed based on unstratified log-rank  without any care given to loss of power due to stratification. A much desired strategy would be to assess the loss of power due to stratification in the design stage and should compensate for that by increasing the target number of events to avoid designing an under-powered study. Our simulation results indicate that HR estimates from multivariate Cox analysis are relatively more accurate and precise  compared to stratified Cox analysis even when  the data was not generated from multivariate Cox model.  This finding goes quite well with the present understanding that stratification compromises with the stability of estimate of HR. The  possible explanation for the smaller bias  with multivariate Cox analysis even under ``stratified Cox model'' is that multivariate Cox model can adapt to the data better so that it can produce a estimate closer the truth.  Our simulation is not exhaustive (e.g., it did not consider the prognostic effect on treatment to control HR) and findings are not absolute, however, it raises serious point about the necessity of evaluating the impact of stratification factors in time to event analyses.





\end{document}